\documentclass[reprint,superscriptaddress,aps,prl,floatfix,nofootinbib,bibnotes]{revtex4-1}
\usepackage[colorlinks=true,linkcolor=black,citecolor=blue, urlcolor=blue,bookmarks=false]{hyperref}
\hypersetup{breaklinks=true}
\usepackage[utf8]{inputenc}
\usepackage{natbib,slashed}
\usepackage{bm}
\usepackage{dcolumn}
\usepackage{array}
\usepackage{amsmath,mathtools}
\usepackage{amssymb}
\usepackage{multirow}
\usepackage{graphicx,subcaption}
\usepackage{tensor}
\usepackage{comment}
\usepackage{enumitem}
\usepackage{float}
\usepackage[dvipsnames]{xcolor}
\usepackage[normalem]{ulem}  
\usepackage{color} 

\renewcommand{\sout}{\bgroup \color{red} \ULdepth=-.5ex \ULset}

\newcommand{\vev}[1]{\langle{#1}\rangle}

\newcommand{\go}{g_{\Omega}}
\newcommand{\rot}{\text{rot}}
\newcommand{\vac}{\text{vac}}

\begin{document}
\title{Spin-1 quarkonia in a rotating frame and their spin contents}

\author{HyungJoo Kim}
\email{hugokm0322@gmail.com}
\affiliation{Department of Physics and Institute of Physics and Applied Physics, Yonsei University, Seoul 03722, Korea}

\author{Sungtae Cho}
\email{sungtae.cho@kangwon.ac.kr}
\affiliation{Division of Science Education, Kangwon National University, Chuncheon 24341, Korea}
\author{Su Houng Lee}%
\email{suhoung@yonsei.ac.kr}
\affiliation{Department of Physics and Institute of Physics and Applied Physics, Yonsei University, Seoul 03722, Korea}

\date{\today}
\begin{abstract}
We propose a new way of studying the spin content of a hadron by looking at its response in a rotating frame. By collecting all responses of quarks and gluons in a rotating frame, we describe the spin-rotation coupling of spin-1 quarkonia and thereby reveal their spin contents in a relativistic formalism.
We demonstrate that both the perturbative and non-perturbative contributions  in the operator product expansion follow a universal formula that identifies the spin-rotation coupling with unit strength. 
This allows us to recognize the total spin-1 of the vector and axial vector quarkonia in terms of the total angular momentum of quarks and gluons. Specifically, we find the  spin contents of $J/\psi$, $\chi_{c1}$, $\Upsilon(1S)$, and $\chi_{b1}$  are slightly different from the naive quark model picture. For example, the $J/\psi$ is traditionally considered as an S-wave particle, but we find quarks do not carry all of the total spin.
\end{abstract}


\maketitle
The reaction of particles with spin in a rotating frame has been of interest for a 
long time since the mid 1910s when it was realized that mechanical rotation can polarize the particle spins through the Barnett effect\cite{barnett1915}. This effect is understood by spin-rotation coupling and recently with the development of experimental techniques led to new wave of active research field such as measurement of the spin polarization of hadrons in heavy-ion collisions. Following the measurements of the global spin polarization of $\Lambda$ hyperons\cite{STAR:2017ckg}, the global spin alignments of vector mesons such as $K^{*0}$, $\phi$, and $J/\psi$ have been measured recently by ALICE and STAR collaborations\cite{ALICE:2019aid,STAR:2022fan,ALICE:2022sli}. While the hyperon polarization phenomena are consistent with  theoretical works\cite{Liang:2004ph,Betz:2007kg,Becattini:2007sr,Becattini:2007nd,STAR:2007ccu,STAR:2018gyt}, the various aspects of the measured vector meson spin alignments are still not clearly understood. As the medium in a heavy-ion collision will evolve through different phases, the overall spin polarization will probably depend on both the total spin and the spin content that varies for different hadrons. Therefore a quantitative analysis of the spin contents of hadrons is required. From a theoretical perspective, however, this has not been fully achieved so far because of the non-perturbative aspects of the strong interaction.

In this work, we propose a new method to study the spin content of a hadron by looking at its response in a rotating frame.
Because hadrons are composite particles, their spin-rotation coupling should be described by  the responses of quarks and gluons in a rotating frame. 
This therefore implies that by studying the spin-rotation coupling of a hadron, we can understand its spin content, which is one of the fundamental questions of nuclear physics especially in the future EIC(Electron-Ion Collider) project\cite{Accardi:2012qut}.  Therefore, as a first step we implement this idea to study the spin contents of spin-1 quarkonia.

In order to do so, let us first consider two reference frames: an inertial frame and a (non-inertial) rotating frame which rotates with an angular velocity $\vec{\Omega}$ with respect to the inertial frame. Then, for a classical particle, it is easy to verify that  the Hamiltonian in the inertial frame($H_i$) and the Hamiltonian in the rotating frame($H_r$) are related by $H_r=H_i-\vec{L}\cdot \vec{\Omega}$ where $\vec{L}$ is the orbital angular momentum of the particle. 
For a particle with intrinsic spin, it seems natural to generalize this relation to $H_r=H_i-(\vec{L}+\vec{S})\cdot \vec{\Omega}$. The appearance of the spin-rotation coupling was proposed by Mashhoon\cite{Mashhoon:1988zz} for particles of any spin but the idea  relied on special relativity and hence  warrants a more general derivation based on general relativity. Indeed, for spin-1/2 Dirac particles, the spin-rotation coupling can be explicitly derived from the Dirac equation in a rotating frame,
\begin{align}
    [i\gamma^\mu D_\mu-m+ \gamma^0 (
    \hat{L}_q+\hat{S}_q)\cdot \vec{\Omega} ]\Psi=0,\label{Diraceq}
\end{align}
where $D_\mu=\partial_\mu+igA_\mu$ is the covariant derivative and $\hat{L}_q=\vec{x}\times (-i\vec{D})$ and $\hat{S}_q=\frac{1}{2}\gamma^0\vec{\gamma}^{\,}\gamma^5$ are the orbital and spin angular momentum operator for Dirac fields, respectively. Expressing this equation in the form $i\partial_0 \Psi=H\Psi$, we indeed find $H_r=H_i-(\vec{L}+\vec{S})\cdot \vec{\Omega}$ for Dirac particles; see  \cite{deOliveira:1962apw,Hehl:1990nf,de1962representations,huang1994dirac,Papini:2002cp,Cai:1991wj,ryder1998relativistic,ryder2008spin} for more details. 

On the other hand, for massive spin-1 particles, only  limited discussions were given in \cite{mashhoon1989electrodynamics,Cai:1991wj}.
Recently, Kapusta \textit{et al.} investigated the spin-rotation coupling for massive spin-1 vector particles starting from the Proca equation in a rotating frame\cite{Kapusta:2020dco}. However, they found that the Hamiltonian in the non-relativistic limit is reduced to   $H_r=H_i-(\vec{L}+\frac{1}{2} \vec{S})\cdot\vec{\Omega}$ at the leading order in $\vec{\Omega}$(or vorticity), which is not consistent with our natural expectations. Hence, it is imperative   to derive the strength of  the spin-rotation coupling for spin-1 particles especially for vector mesons in a model independent way based on Quantum Chromodynamics(QCD).

In this work, we introduce a free parameter $g_\Omega$, so-called gravitomagnetic moment in \cite{Buzzegoli:2021jeh}, which represents the strength of the spin-rotation coupling for spin-1 system composed of a heavy quark and its anti-quark,
\begin{align}
    H_r=H_i-\go \vec{S}\cdot\vec{\Omega}.
\end{align}
To derive the value of $\go$ on the basis of quarks and gluons degrees of freedom, let us first consider a two-point correlation function for the vector current, 
\begin{align}
    \Pi^{\mu\nu}(q)=i\int d^4x e^{iqx} \big\langle 0| \textrm{T}\big[j^\mu (x) j^\nu  (0)\big]|0\big\rangle.\label{correlationftn}
\end{align} 
Without loss of generality, we put the system at the center of the rotation and pick out a right circularly polarized state with $\epsilon_\mu^+=(0,1,i,0)/\sqrt{2}$ in the rotating frame in which the polarization axis and the angular velocity are along the same $z$-direction, i.e. $q_{\mu}=(\omega,0)$ and $\vec{\Omega}=(0,0,\Omega)$. Under this circumstance, the orbital motion of the system is absent at the center. Furthermore, the angular velocity is assumed to be small enough so that rotation effects can be expanded in terms of $\Omega$ by perturbation theory. Because we are mainly interested in the terms linear in $\Omega$, we define the relevant component of the correlation function as 
\begin{align}
\Pi^+(\omega)=\omega^2 \Pi^{\rm{vac}}(\omega^2)+\omega\Omega\Pi^{\rm{rot}}(\omega^2),\label{piplus}
\end{align}
where $\Pi^+(\omega)=\Pi^{\mu\nu}(\omega,0)\epsilon_\mu^+ \epsilon_\nu^{+*}$. $\Pi^{\rm{vac}}(\omega^2)$ 
is the vacuum invariant function which is used to study vacuum properties of hadrons and $\Pi^{\rm{rot}}(\omega^2)$ is a new function appearing in the rotating frame. Once we obtain $\Pi^{\rm{rot}}(\omega^2)$, $g_\Omega$ can be extracted by comparing it with $\Pi^{\rm{vac}}(\omega^2)$. 

To this end, let us first discuss the phenomenological structure of $\Pi^+(\omega)$. In an inertial frame, $\Omega$ dependent terms do not appear so that $\Pi^+(\omega)=\omega^2 \Pi^{\text{vac}}(\omega^2)$. Once we turn on the rotation, the energy of a right circularly polarized state is shifted by $-\go\Omega$ as we assumed. If this energy shift happens equally, i.e. $\go$ is universal, for all vector states contributing in Eq.\eqref{correlationftn},  we can expect that $\Pi^+(\omega)\to\Pi^+(\omega+\go\Omega)$ in the rotating frame. 
By comparing this with Eq.(\ref{piplus}), we can infer a simple relation between $\Pi^\vac(\omega^2)$ and $\Pi^\rot(\omega^2)$,
\begin{align}
    \Pi^\rot_{\text{phen}}(\omega^2)=2\go\Big\lbrace \Pi^{\rm{vac}}(\omega^2)+\omega^2\frac{\partial\Pi^{\rm{vac}}(\omega^2)}{\partial \omega^2}\Big\rbrace,\label{simplerelation}
\end{align}
where the subscript `phen' indicates that this relation is derived from a phenomenological point of view.
Furthermore, by applying this relation to the  dispersion relation in vacuum, we obtain the following useful expression
\begin{align}
    \Pi^{\rm{rot}}_{\text{phen}}(\omega^2)=\frac{\go}{\pi}\int^\infty_{4m^2}ds\frac{2s\mathrm{Im}\Pi^{\rm{vac}}(s)}{(s-\omega^2)^2}.\label{simplerelationds}
\end{align}

In the following, we will show that the Operator Product Expansion(OPE) satisfies a universal formula
where the 
`rot' part of the perturbative and non-perturbative contributions 
are related to their respective  `vac' parts  as in Eq.\eqref{simplerelation} but with $\go=1$. To compute the OPE of Eq.\eqref{correlationftn} in the rotating frame, we first consider quark propagators in the rotating frame. 
Referring to Eq.(\ref{Diraceq}), we use the following expansion of the quark propagator in terms of the coupling constant and $\Omega$,
\begin{align}
    S(x&,0)=S^{(0)}(x)  +\sum^\infty_{n=1}(-1)^n\int dz_1\ldots dz_n S^{(0)}(x-z_1)\nonumber\\
    & \times \big[\Delta\mathcal{I}(z_1)\big]S^{(0)}(z_1-z_2)\ldots\big[\Delta\mathcal{I}(z_n)\big]S^{(0)}(z_n),\label{proppert}
\end{align}
where $S^{(0)}(x)$ is the free quark propagator and 
$\Delta\mathcal{I}=g\slashed{A}+\gamma^0 ((\hat{L}_{q})_z+(\hat{S}_{q})_z)\Omega$ includes all the interaction terms.
For later convenience, we also distinguish the orbital angular momentum operator into two pieces, $\hat{L}_q=\hat{L}_k+\hat{L}_p$ where $\hat{L}_k=\vec{x}\times \vec{p}$ is the kinetic part and $\hat{L}_p=\vec{x}\times (-g\vec{A}(x))$ is the potential part, respectively. Furthermore, gluon fields appearing in the interaction terms are also modified by the rotation. Within Fock-Schwinger gauge($x^\mu A_\mu(x)=0$), the gluon field in the rotating frame is expressed as
\begin{align}
    A_\mu(x)=&-\frac{1}{2}x^\nu G_{\mu\nu}(0)\nonumber\\
    &-\frac{1}{3}x^\nu x^\alpha (\Gamma^\rho_{\alpha\mu}G_{\rho\nu}(0)+\Gamma^\rho_{\alpha\nu}G_{\mu \rho}(0))+\ldots,\label{gluonrot}
\end{align}
where $\Gamma^\rho_{\mu\nu}$ denotes the Christoffel symbol: only $\Gamma^2_{01}=\Omega$ and $\Gamma^1_{02}=-\Omega$ contribute in this work. Throughout this work, we denote the contribution of $\Omega$ linear terms in Eq.\eqref{gluonrot} as $\hat{J}_g$ because we later found that the sum gives the same contribution with the gluon's total angular momentum operator, $\hat{J}_g=\vec{x}\times (\vec{E}\times \vec{B})$, which can be computed by the method discussed in \cite{Balitsky:1997rs}. By collecting terms linear in $\Omega$ from Eq.(\ref{proppert}), we compute the OPE for $\Pi^{\rm{rot}}$ upto operators of dimension 4,
\begin{align}
    \Pi^{\rot}_{\text{OPE}}(Q^2)=\sum_{\substack{i=S_q,L_k,L_p,J_g}} \Pi^{\rot}_{\text{I},i}(Q^2)+
\Pi^\rot_{G_0,i}(Q^2),
\end{align}
where $\Pi_{\text{I},i}^{\rot}$ denotes the leading perturbative part and  $\Pi^{\text{rot}}_{G_0 ,i}=C_i(Q^2) \cdot G_0$ denotes the leading non-perturbative part with $C_i(Q^2)$ being the  Wilson coefficients for the scalar gluon condensates $G_0\equiv\vev{\frac{\alpha_s}{\pi} G^a_{\mu\nu}G^{a,\mu\nu}}=(0.35\, \text{GeV})^4$.
\begin{figure}[ht]
\centering
\includegraphics[width=0.4\linewidth]{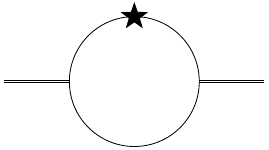}
\caption{The leading perturbative diagram}
\label{leadingpert}
\end{figure}

The leading perturbative diagram which contributes to $\Pi_{\rm{I}}^{\rm{rot}}$  is shown in Fig.\ref{leadingpert}. Here, the propagator with one star symbol($\bigstar$) indicates the insertion of interaction terms linear in $\Omega$ on the quark propagator. The perturbative result can be expressed by a simple dispersion relation,
\begin{align}
    \Pi^{\rm{rot}}_{\rm{I}}(Q^2)&=\frac{1}{\pi}\int^\infty_{4m^2} ds \frac{\mathrm{Im}\Pi^{\rm{rot}}_{\text{I}}(s)}{s+Q^2},
\end{align}
where $\Pi^{\text{rot}}_{\text{I}}(s)=\Pi^{\rm{rot}}_{\text{I},S_q}(s)+\Pi^{\rm{rot}}_{\text{I},{L_k}}(s)$ and
\begin{align}
    \mathrm{Im}\Pi^{\rm{rot}}_{\text{I},S_q}&(s)=\frac{3m^2}{2\pi\sqrt{s(s-4m^2)}},\\
    \mathrm{Im}\Pi^{\rm{rot}}_{\text{I},{L_k}}&(s)=\frac{(s-m^2)\sqrt{s(s-4m^2)}}{2\pi s^2}.
\end{align}
Here, $\hat{L}_p$ and $\hat{J}_g$ have no contribution because they already contain gluon fields. Now we can extract $\go$ in the perturbative region by comparing this OPE result with the corresponding phenomenological expression given in Eq.(\ref{simplerelationds}),
\begin{align}
    \frac{1}{\pi}\int^\infty_{4m^2} ds \frac{\mathrm{Im}\Pi^{\rm{rot}}_{\text{I}}(s)}{s+Q^2}=\frac{\go}{\pi}\int^\infty_{4m^2}ds\frac{2s\mathrm{Im}\Pi^\vac_\text{I}(s)}{(s+Q^2)^2},\label{leadingrelation}
\end{align}
where $\Pi^\vac_\text{I}(s)$ is the leading perturbative contribution for the vacuum OPE which is given by $u(3-u^2)/(8\pi)$ with $u=\sqrt{1-4m^2/s}$. Here, the left hand side is directly computed by collecting all responses of the quarks in the rotating frame and the right hand side is derived from a phenomenological point of view. Interestingly, we find
\begin{align}
   \text{Im} \Pi^\rot_{\text{I}}(s)=2\Big\lbrace \text{Im}\Pi^{\rm{vac}}_{\text{I}}(s)+s\frac{\partial \text{Im}\Pi^{\rm{vac}}_{\text{I}}(s)}{\partial s} \Big\rbrace.
\end{align}
Therefore, by integration by parts of the right hand side, we can explicitly show that $\go=1$ in the leading perturbative region. The leading perturbative diagram describes the free quark system so this finding indicates that when free quarks form a spin-1 vector state, their total energy shift is just given by the sum of the energy shift of each quark in the rotating frame. It should be noted that this goes beyond the non-relativistic quark model as both the spin and orbital angular momentum of quarks are taken into account relativistically.

\begin{figure}[ht]
\centering
\includegraphics[width=\linewidth]{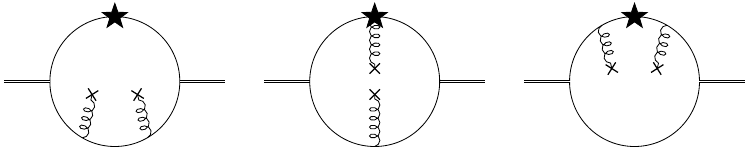}
\caption{The leading non-perturbative diagrams}
\label{spinrotation}
\end{figure}
Now let us examine the non-perturbative region of the  OPE. The leading non-perturbative gluon contribution comes from diagrams shown in  Fig.\ref{spinrotation}.  Here, each curly line indicates one background gluon field inside the propagator  in Eq.(\ref{proppert}). The OPE result is given as $\Pi^\rot_{G_0}=(C_{S_q}+C_{L_k}+C_{L_p}+C_{J_g})G_0$ and the Wilson coefficients are expressed using $J_N=\int^1_0 dx [1+x(1-x)Q^2/m^2]^{-N}$ as follows.
\begin{align}
C_{S_q}&=-\frac{2-y+12 J_2-26 J_3+12 J_4}{12 Q^4},\\
C_{L_k}&=\frac{11-4y-(8+y)J_1+13 J_2-16 J_3}{72Q^4},\\
C_{L_p}&=-\frac{13+2y-(12+5y) J_1+3 J_2-4 J_3}{72 Q^4},\\
C_{J_g}&=\frac{13-(2+2y) J_1-23 J_2+12 J_3}{36 Q^4}.
\end{align}
By comparing this result with the counterpart of the vacuum OPE, we indeed find 
\begin{align}
    \Pi^{\rm{rot}}_{G_0}(Q^2)= 2\Big\lbrace \Pi^{\rm{vac}}_{G_0}(Q^2)+Q^2\frac{\partial\Pi^{\rm{vac}}_{G_0}(Q^2)}{\partial Q^2}\Big\rbrace,
    \label{simpleOPE}
\end{align}
where $\Pi^\vac_{G_0}=\frac{1}{12Q^4}(-1+3J_2-2J_3)G_0$\cite{Shifman:1978bx}. Therefore, $\go=1$ even  in the non-perturbative region. 

Let us think about the implication of our finding. In fact, what we have computed in the OPE is equivalent to the expectation value of the total angular momentum operator in QCD, $\vec{J}_{\text{QCD}}=\int d^3x (\frac{1}{2}\bar{\psi}\vec{\gamma} \gamma_5 \psi+\psi^{\dagger} ( \vec{x}\times (-i\vec{D}))\psi+\vec{x}\times (\vec{E}\times \vec{B}))$\cite{Ji:1996ek}, with respect to a right circularly polarized vector current. Similarly, what we have considered in the phenomenological side is nothing else but the expectation value of $\go \vec{S}\cdot\vec{\Omega}$ where $\vec{S}$ is the spin-1 operator.
Therefore, we can conclude that $\go=\vev{ (\vec{J}_{\text{QCD}})_z}/\vev{(\vec{S}\,)_z}$ where $\vev{\cdots}=i\int d^4 x e^{iqx} \epsilon^+_\mu \epsilon^{+*}_\nu\vev{0|\textrm{T}\big[j^\mu(x) \cdots j^\nu(0)\big]|0}$. 
Then our finding, i.e. $\go=1$, just shows that the total spin of the system is equal to the total angular momentum of quarks and gluons. Even though we have checked only the leading perturbative and non-perturbative diagrams in the vector channel, this should be valid at any order of Feynman diagrams and also in any other system because of the angular momentum conservation. We also compute OPE for the axial vector case and indeed observe it satisfies the same property as in the vector case. The OPE results for the axial vector are listed below.
\begin{align}
\mathrm{Im}\Pi^{\rm{rot}}_{\text{I},{S_q}}&(s)=\frac{3m^2\sqrt{s(s-4m^2)}}{2\pi s^2},\\
\mathrm{Im}\Pi^{\rm{rot}}_{\text{I},{L_k}}&(s)=\frac{(s-m^2)\sqrt{s(s-4m^2)}}{2\pi s^2},\\
C_{S_q}&=-\frac{6-y-12 J_2+6 J_3}{12 Q^4},\\
C_{L_k}&=\frac{7-4y-(8+y)J_1+25 J_2-24 J_3}{72Q^4},\\
C_{L_p}&=-\frac{9+2y+(4-5y) J_1-17 J_2+4 J_3}{72 Q^4},\\
C_{J_g}&=\frac{1+(6-2y) J_1-3 J_2-4 J_3}{36 Q^4}.
\end{align}

The above findings can be utilized to calculate the spin contents of spin-1 quarkonia such as $J/\psi$, $\Upsilon$(1S), $\chi_{c1}$, and $\chi_{b1}$ using QCD sum rules. Although $\go$ should be 1 even for bound states, we are particularly interested in the fraction of $\go$ carried by each angular momentum operator($\hat{S}_q$, $\hat{L}_k$, $\hat{L}_p$, $\hat{J}_g$). To this end, let us recall the phenomenological expression once again. In Eq.(\ref{simplerelationds}), $\text{Im}\Pi^\vac(s)$ represents the spectral density in which all physical states that can couple to the vector or axial vector current are involved. In order to pick out only the ground state, it is often modeled to have a single ground state pole with the continuum that is described by the perturbative diagrams,
\begin{align}
    \text{Im}\Pi^\vac(s)=\pi f_0\delta(s-m_0^2)+\theta(s-s_0)\text{Im}\Pi^\vac_\text{I}(s),
\end{align}
where $f_0$ denotes the coupling strength between the current and the ground state, $m_0$ is the ground state mass, $s_0$ is the threshold for the leading perturbative continuum. Then the phenomenological side is expressed as
\begin{align}
    \Pi^{\rm{rot}}_{\text{phen}}(Q^2)=\frac{2\go f_0}{(Q^2+m_0^2)^2}+\frac{1}{\pi}\int^\infty_{s_0}ds\frac{\mathrm{Im}\Pi^{\rot}_{\text{I}}(s)}{(s+Q^2)},
\end{align}
where only the $\go$ for the ground state remains and the continuum part is re-expressed in terms of $\Pi^\rot_\text{I}$ using Eq.(\ref{leadingrelation}).
By connecting this expression to the OPE side and performing Borel transformation, we can obtain a new sum rule for $\Pi^\rot$ as
\begin{align}
\bar{\mathcal{M}}^{\rot}&\equiv \mathcal{B}\big[\Pi^{\rot}_{\text{OPE}}(Q^2)\big]-\int^\infty_{s_0}ds e^{-s/M^2}\mathrm{Im}\Pi^{\rot}_{\text{I}}(s)\nonumber\\
&=\frac{2\pi\go f_0 m_0^2}{M^2}e^{-m_0^2/M^2}.
\end{align}
Here, Borel transformation is defined by 
\begin{align}
 \mathcal{B}\equiv\lim_{\substack{Q^2/n \rightarrow M^2, \\ n,Q^2
  \rightarrow \infty}}
  \frac{\pi(Q^2)^{n+1}}{n!}\left(-\frac{d}{dQ^2}\right)^n,
\end{align}
where the Borel mass, $M$, is an auxiliary parameter that simultaneously controls convergence of the OPE series and dominance of the ground state contribution in the sum rule.
Similarly, the vacuum sum rule is expressed as
\begin{align}
\bar{\mathcal{M}}^{\rm{vac}}&\equiv\mathcal{B}\big[\Pi^\vac_{\text{OPE}}(Q^2)\big]-\int^\infty_{s_0}ds e^{-s/M^2}\mathrm{Im}\Pi^{\rm{vac}}_{\text{I}}(s) \nonumber\\
&=\pi f_0 e^{-m_0^2/M^2}.\label{borelsumrulevac}
\end{align}
Then the fraction of $\go$ for the ground state can be extracted from the following equation,
\begin{align}
    \go(M,s_0)=-\frac{M^2}{2}\frac{\bar{\mathcal{M}}^{\rm{rot}}}{\partial\bar{\mathcal{M}}^{\rm{vac}}/\partial(1/M^2)},\label{go}
\end{align}
by decomposing $\bar{\mathcal{M}}^{\rm{rot}}$ into contributions from $\hat{S}_q$, $\hat{L}_k$, $\hat{L}_p$, and $\hat{J}_g$. This fraction indicates the spin content of the ground state as it represents the contribution of each angular momentum to the total spin.

In an actual analysis, we need to specify an effective threshold($\bar{s}_0$) and a reliable range of Borel mass, so-called Borel window, because we truncated the OPE series at the order of dimension 4 operator and simplified the spectral function. 
In this work, we will simply employ the values of $\bar{s}_0$ and Borel window used in the conventional analysis for determining the $m_0$ through the vacuum sum rule in the presence of $\alpha_s$-correction\cite{Reinders:1984sr}.
The ground state mass, $m_0$, can also be extracted from Eq.\eqref{borelsumrulevac} as a function of $M$ and $s_0$. We then determine $\bar{s}_0$ by tuning $s_0$ to make $m_0(M,s_0)$ least sensitive to $M$ within the Borel window. Here, we determine $M_{min}$ by setting the condition that the sum of $\alpha_s$-correction and $G_0$ contribution should not exceed 30$\%$ of the total vacuum OPE, and $M_{max}$ by setting the condition that the ground state contribution should be greater than 60$\%$ of the total OPE; See \cite{Morita:2009qk,Gubler:2018ctz} for more details on QCD sum rules.  For input parameters, we use $m_c(p^2=-m_c^2)=1.262\,\text{GeV}$, $\alpha_s(8m_c^2)=0.21$ for charmonia and $m_b(p^2=-m_b^2)=4.12\,\text{GeV}$, $\alpha_s(8m_b^2)=0.158$ for bottomonia following \cite{Morita:2009qk}. The values of $\bar{s}_0$ and Borel window used in this work are listed in Table \ref{thresholdborelwindow}. 
\begin{table}[h!]
    \centering
        \caption{$\sqrt{\bar{s}_0}$ and Borel window for spin-1 quarkonia} 
{\renewcommand{\arraystretch}{1.2}   
    \begin{tabular}{c|>{\centering}p{1.2cm}>{\centering}p{1.2cm}>{\centering}p{1.2cm}>{\centering\arraybackslash}p{1.2cm}} 
    \hline\hline
         &  $J/\psi$ & $\chi_{c1}$ & $\Upsilon$(1S)& $\chi_{b1}$\\ \hline
         $\sqrt{\bar{s}_0}$ [GeV] &3.5& 4.0& 10.3& 11\\ 
         ($M_{\text{min}}$,$M_{\text{max}}$) [GeV]& (1,2.3) &(1.4,2.3) &(3,5.5) &(3.6,4.9) \\ \hline\hline
    \end{tabular}}
    \label{thresholdborelwindow}
\end{table}

Then we finally estimate the spin contents of the spin-1 quarkonia by averaging the contribution of each angular momentum operator in $\go(M,\bar{s}_0)$ over the given Borel window. We also calculate the variance of each contribution to estimate the uncertainty. The average values and their uncertainties(subscript) are listed in Table \ref{spincontent}. Here, it should be noted that while the sum of $\hat{L}_k$ and $\hat{L}_p$ are gauge invariant, each of them is gauge dependent. Thus their values should be understood within Fock-Schwinger gauge.
\begin{table}[h!]
    \centering
        \caption{The spin contents of spin-1 quarkonia} 
{\renewcommand{\arraystretch}{1.2}   
    \begin{tabular}{>{\centering}p{0.6cm}|>{\centering}p{1.7cm}>{\centering}p{1.7cm}|>{\centering}p{1.7cm}>{\centering\arraybackslash}p{1.7cm}} 
    \hline\hline
    &\multicolumn{2}{c|}{Vector} & \multicolumn{2}{c}{Axial vector} \\ \cline{2-5}
         &  $J/\psi$ &$\Upsilon$(1S)&  $\chi_{c1}$ & $\chi_{b1}$\\ \hline
         $S_q$ &$0.88_{1.8\text{e-}4}$& $0.92_{7.6\text{e-}5}$& $0.40_{8.2\text{e-}5}$& $0.43_{1.1\text{e-}5}$\\ 
         $L_k$& $ 0.11_{4.9\text{e-}4}$& $0.076_{7.8\text{e-}5}$& $0.61_{5.8\text{e-}6}$&$0.57_{1.0\text{e-}5}$ \\ 
         $L_p$ &$2.0\text{e-}3_{2.9\text{e-}6}$ &$3.5\text{e-}5_{3.0\text{e-}10}$ &$8.2\text{e-}4_{2.3\text{e-}8}$&$\textbf{--}1.0\text{e-}5_{3.4\text{e-}10}$ \\
         $J_g$ & $8.0\text{e-}3_{5.9\text{e-}5}$& $1.5\text{e-}4_{7.3\text{e-}9}$&$\textbf{--} 0.015_{5.2\text{e-}5}$&$\textbf{--} 5.2\text{e-}5_{2.3\text{e-}8}$  \\ \hline\hline
    \end{tabular}}
    \label{spincontent}
\end{table}

In all cases the sum of the four components is exactly 1 as we expected, but the spin contents are quite different from each other. While the quark spin has the most dominant contribution for the vector quarkonia, the kinetic part of the orbital angular momentum has larger contribution for the axial vector quarkonia. The bottomonia results are somewhat comparable with the non-relativistic quark model picture but the spin contents start to deviate from this picture as the quark mass becomes lighter due to the significant increase in  the contribution of $\hat{L}_k$. For example, the $J/\psi$ is traditionally considered as an S-wave particle but now we  find that the quark spin does not carry all of the total spin as in the case of the proton spin\cite{EuropeanMuon:1987isl}. Furthermore, the overall contributions of $\hat{L}_p$ and $\hat{J}_g$ are quite small but indispensable to make $\go=1$ exactly.
 In the perturbative region, $\hat{L}_p$ and $\hat{J}_g$ start to contribute at the next-to-leading order diagrams, so more accurate analysis requires the computation of $\alpha_s$-correction for $\Pi^\rot$.

In conclusion, 
we have proven that $\go=1$ for heavy composite particles with spin-1 and simultaneously have identified how the angular momenta of quarks and gluons add up to their total spin in a relativistic way.
The most crucial finding in this work  is the discovery of the universal formula in the OPE which is given by a simple relation between the rotating frame part and the corresponding inertial frame part. 
Because this relation indicates that the total spin of the system is equal to the total angular momentum of its constituents, the method discussed in this work  can be applied to any other systems to elucidate their spin contents. Therefore, we have taken the first step toward understanding the spin contents of more complex hadrons, especially for nucleons, using the spin-rotating coupling.

\section*{Acknowledgements}
This work was supported by Samsung Science and Technology
Foundation under Project Number SSTF-BA1901-04, and by the Korea National Research Foundation under the grant number No.2020R1F1A1075963 and No.2019R1A2C1087107.

\bibliographystyle{apsrev4-1}
\bibliography{refs}

\providecommand{\noopsort}[1]{}\providecommand{\singleletter}[1]{#1}%
\begin{thebibliography}{31}%
\makeatletter
\providecommand \@ifxundefined [1]{%
 \@ifx{#1\undefined}
}%
\providecommand \@ifnum [1]{%
 \ifnum #1\expandafter \@firstoftwo
 \else \expandafter \@secondoftwo
 \fi
}%
\providecommand \@ifx [1]{%
 \ifx #1\expandafter \@firstoftwo
 \else \expandafter \@secondoftwo
 \fi
}%
\providecommand \natexlab [1]{#1}%
\providecommand \enquote  [1]{``#1''}%
\providecommand \bibnamefont  [1]{#1}%
\providecommand \bibfnamefont [1]{#1}%
\providecommand \citenamefont [1]{#1}%
\providecommand \href@noop [0]{\@secondoftwo}%
\providecommand \href [0]{\begingroup \@sanitize@url \@href}%
\providecommand \@href[1]{\@@startlink{#1}\@@href}%
\providecommand \@@href[1]{\endgroup#1\@@endlink}%
\providecommand \@sanitize@url [0]{\catcode `\\12\catcode `\$12\catcode
  `\&12\catcode `\#12\catcode `\^12\catcode `\_12\catcode `\%12\relax}%
\providecommand \@@startlink[1]{}%
\providecommand \@@endlink[0]{}%
\providecommand \url  [0]{\begingroup\@sanitize@url \@url }%
\providecommand \@url [1]{\endgroup\@href {#1}{\urlprefix }}%
\providecommand \urlprefix  [0]{URL }%
\providecommand \Eprint [0]{\href }%
\providecommand \doibase [0]{http://dx.doi.org/}%
\providecommand \selectlanguage [0]{\@gobble}%
\providecommand \bibinfo  [0]{\@secondoftwo}%
\providecommand \bibfield  [0]{\@secondoftwo}%
\providecommand \translation [1]{[#1]}%
\providecommand \BibitemOpen [0]{}%
\providecommand \bibitemStop [0]{}%
\providecommand \bibitemNoStop [0]{.\EOS\space}%
\providecommand \EOS [0]{\spacefactor3000\relax}%
\providecommand \BibitemShut  [1]{\csname bibitem#1\endcsname}%
\let\auto@bib@innerbib\@empty
\bibitem [{\citenamefont {Barnett}(1915)}]{barnett1915}%
  \BibitemOpen
  \bibfield  {author} {\bibinfo {author} {\bibfnamefont {S.~J.}\ \bibnamefont
  {Barnett}},\ }\href@noop {} {\bibfield  {journal} {\bibinfo  {journal}
  {Physical review}\ }\textbf {\bibinfo {volume} {6}},\ \bibinfo {pages} {239}
  (\bibinfo {year} {1915})}\BibitemShut {NoStop}%
\bibitem [{\citenamefont {Adamczyk}\ \emph {et~al.}(2017)\citenamefont
  {Adamczyk} \emph {et~al.}}]{STAR:2017ckg}%
  \BibitemOpen
  \bibfield  {author} {\bibinfo {author} {\bibfnamefont {L.}~\bibnamefont
  {Adamczyk}} \emph {et~al.} (\bibinfo {collaboration} {STAR}),\ }\href
  {\doibase 10.1038/nature23004} {\bibfield  {journal} {\bibinfo  {journal}
  {Nature}\ }\textbf {\bibinfo {volume} {548}},\ \bibinfo {pages} {62}
  (\bibinfo {year} {2017})},\ \Eprint {http://arxiv.org/abs/1701.06657}
  {arXiv:1701.06657 [nucl-ex]} \BibitemShut {NoStop}%
\bibitem [{\citenamefont {Acharya}\ \emph {et~al.}(2020)\citenamefont {Acharya}
  \emph {et~al.}}]{ALICE:2019aid}%
  \BibitemOpen
  \bibfield  {author} {\bibinfo {author} {\bibfnamefont {S.}~\bibnamefont
  {Acharya}} \emph {et~al.} (\bibinfo {collaboration} {ALICE}),\ }\href
  {\doibase 10.1103/PhysRevLett.125.012301} {\bibfield  {journal} {\bibinfo
  {journal} {Phys. Rev. Lett.}\ }\textbf {\bibinfo {volume} {125}},\ \bibinfo
  {pages} {012301} (\bibinfo {year} {2020})},\ \Eprint
  {http://arxiv.org/abs/1910.14408} {arXiv:1910.14408 [nucl-ex]} \BibitemShut
  {NoStop}%
\bibitem [{\citenamefont {Abdallah}\ \emph {et~al.}(2022)\citenamefont
  {Abdallah} \emph {et~al.}}]{STAR:2022fan}%
  \BibitemOpen
  \bibfield  {author} {\bibinfo {author} {\bibfnamefont {M.}~\bibnamefont
  {Abdallah}} \emph {et~al.} (\bibinfo {collaboration} {STAR}),\ }\href@noop {}
  {\  (\bibinfo {year} {2022})},\ \Eprint {http://arxiv.org/abs/2204.02302}
  {arXiv:2204.02302 [hep-ph]} \BibitemShut {NoStop}%
\bibitem [{\citenamefont {Acharya}\ \emph {et~al.}(2022)\citenamefont {Acharya}
  \emph {et~al.}}]{ALICE:2022sli}%
  \BibitemOpen
  \bibfield  {author} {\bibinfo {author} {\bibfnamefont {S.}~\bibnamefont
  {Acharya}} \emph {et~al.} (\bibinfo {collaboration} {ALICE}),\ }\href@noop {}
  {\  (\bibinfo {year} {2022})},\ \Eprint {http://arxiv.org/abs/2204.10171}
  {arXiv:2204.10171 [nucl-ex]} \BibitemShut {NoStop}%
\bibitem [{\citenamefont {Liang}\ and\ \citenamefont
  {Wang}(2005)}]{Liang:2004ph}%
  \BibitemOpen
  \bibfield  {author} {\bibinfo {author} {\bibfnamefont {Z.-T.}\ \bibnamefont
  {Liang}}\ and\ \bibinfo {author} {\bibfnamefont {X.-N.}\ \bibnamefont
  {Wang}},\ }\href {\doibase 10.1103/PhysRevLett.94.102301} {\bibfield
  {journal} {\bibinfo  {journal} {Phys. Rev. Lett.}\ }\textbf {\bibinfo
  {volume} {94}},\ \bibinfo {pages} {102301} (\bibinfo {year} {2005})},\
  \bibinfo {note} {[Erratum: Phys.Rev.Lett. 96, 039901 (2006)]},\ \Eprint
  {http://arxiv.org/abs/nucl-th/0410079} {arXiv:nucl-th/0410079} \BibitemShut
  {NoStop}%
\bibitem [{\citenamefont {Betz}\ \emph {et~al.}(2007)\citenamefont {Betz},
  \citenamefont {Gyulassy},\ and\ \citenamefont {Torrieri}}]{Betz:2007kg}%
  \BibitemOpen
  \bibfield  {author} {\bibinfo {author} {\bibfnamefont {B.}~\bibnamefont
  {Betz}}, \bibinfo {author} {\bibfnamefont {M.}~\bibnamefont {Gyulassy}}, \
  and\ \bibinfo {author} {\bibfnamefont {G.}~\bibnamefont {Torrieri}},\ }\href
  {\doibase 10.1103/PhysRevC.76.044901} {\bibfield  {journal} {\bibinfo
  {journal} {Phys. Rev. C}\ }\textbf {\bibinfo {volume} {76}},\ \bibinfo
  {pages} {044901} (\bibinfo {year} {2007})},\ \Eprint
  {http://arxiv.org/abs/0708.0035} {arXiv:0708.0035 [nucl-th]} \BibitemShut
  {NoStop}%
\bibitem [{\citenamefont {Becattini}\ \emph {et~al.}(2008)\citenamefont
  {Becattini}, \citenamefont {Piccinini},\ and\ \citenamefont
  {Rizzo}}]{Becattini:2007sr}%
  \BibitemOpen
  \bibfield  {author} {\bibinfo {author} {\bibfnamefont {F.}~\bibnamefont
  {Becattini}}, \bibinfo {author} {\bibfnamefont {F.}~\bibnamefont
  {Piccinini}}, \ and\ \bibinfo {author} {\bibfnamefont {J.}~\bibnamefont
  {Rizzo}},\ }\href {\doibase 10.1103/PhysRevC.77.024906} {\bibfield  {journal}
  {\bibinfo  {journal} {Phys. Rev. C}\ }\textbf {\bibinfo {volume} {77}},\
  \bibinfo {pages} {024906} (\bibinfo {year} {2008})},\ \Eprint
  {http://arxiv.org/abs/0711.1253} {arXiv:0711.1253 [nucl-th]} \BibitemShut
  {NoStop}%
\bibitem [{\citenamefont {Becattini}\ and\ \citenamefont
  {Piccinini}(2008)}]{Becattini:2007nd}%
  \BibitemOpen
  \bibfield  {author} {\bibinfo {author} {\bibfnamefont {F.}~\bibnamefont
  {Becattini}}\ and\ \bibinfo {author} {\bibfnamefont {F.}~\bibnamefont
  {Piccinini}},\ }\href {\doibase 10.1016/j.aop.2008.01.001} {\bibfield
  {journal} {\bibinfo  {journal} {Annals Phys.}\ }\textbf {\bibinfo {volume}
  {323}},\ \bibinfo {pages} {2452} (\bibinfo {year} {2008})},\ \Eprint
  {http://arxiv.org/abs/0710.5694} {arXiv:0710.5694 [nucl-th]} \BibitemShut
  {NoStop}%
\bibitem [{\citenamefont {Abelev}\ \emph {et~al.}(2007)\citenamefont {Abelev}
  \emph {et~al.}}]{STAR:2007ccu}%
  \BibitemOpen
  \bibfield  {author} {\bibinfo {author} {\bibfnamefont {B.~I.}\ \bibnamefont
  {Abelev}} \emph {et~al.} (\bibinfo {collaboration} {STAR}),\ }\href {\doibase
  10.1103/PhysRevC.76.024915} {\bibfield  {journal} {\bibinfo  {journal} {Phys.
  Rev. C}\ }\textbf {\bibinfo {volume} {76}},\ \bibinfo {pages} {024915}
  (\bibinfo {year} {2007})},\ \bibinfo {note} {[Erratum: Phys.Rev.C 95, 039906
  (2017)]},\ \Eprint {http://arxiv.org/abs/0705.1691} {arXiv:0705.1691
  [nucl-ex]} \BibitemShut {NoStop}%
\bibitem [{\citenamefont {Adam}\ \emph {et~al.}(2018)\citenamefont {Adam} \emph
  {et~al.}}]{STAR:2018gyt}%
  \BibitemOpen
  \bibfield  {author} {\bibinfo {author} {\bibfnamefont {J.}~\bibnamefont
  {Adam}} \emph {et~al.} (\bibinfo {collaboration} {STAR}),\ }\href {\doibase
  10.1103/PhysRevC.98.014910} {\bibfield  {journal} {\bibinfo  {journal} {Phys.
  Rev. C}\ }\textbf {\bibinfo {volume} {98}},\ \bibinfo {pages} {014910}
  (\bibinfo {year} {2018})},\ \Eprint {http://arxiv.org/abs/1805.04400}
  {arXiv:1805.04400 [nucl-ex]} \BibitemShut {NoStop}%
\bibitem [{\citenamefont {Accardi}\ \emph {et~al.}(2016)\citenamefont {Accardi}
  \emph {et~al.}}]{Accardi:2012qut}%
  \BibitemOpen
  \bibfield  {author} {\bibinfo {author} {\bibfnamefont {A.}~\bibnamefont
  {Accardi}} \emph {et~al.},\ }\href {\doibase 10.1140/epja/i2016-16268-9}
  {\bibfield  {journal} {\bibinfo  {journal} {Eur. Phys. J. A}\ }\textbf
  {\bibinfo {volume} {52}},\ \bibinfo {pages} {268} (\bibinfo {year} {2016})},\
  \Eprint {http://arxiv.org/abs/1212.1701} {arXiv:1212.1701 [nucl-ex]}
  \BibitemShut {NoStop}%
\bibitem [{\citenamefont {Mashhoon}(1988)}]{Mashhoon:1988zz}%
  \BibitemOpen
  \bibfield  {author} {\bibinfo {author} {\bibfnamefont {B.}~\bibnamefont
  {Mashhoon}},\ }\href {\doibase 10.1103/PhysRevLett.61.2639} {\bibfield
  {journal} {\bibinfo  {journal} {Phys. Rev. Lett.}\ }\textbf {\bibinfo
  {volume} {61}},\ \bibinfo {pages} {2639} (\bibinfo {year}
  {1988})}\BibitemShut {NoStop}%
\bibitem [{\citenamefont {de~Oliveira}\ and\ \citenamefont
  {Tiomno}(1962)}]{deOliveira:1962apw}%
  \BibitemOpen
  \bibfield  {author} {\bibinfo {author} {\bibfnamefont {C.~G.}\ \bibnamefont
  {de~Oliveira}}\ and\ \bibinfo {author} {\bibfnamefont {J.}~\bibnamefont
  {Tiomno}},\ }\href {\doibase 10.1007/BF02816716} {\bibfield  {journal}
  {\bibinfo  {journal} {Nuovo Cim.}\ }\textbf {\bibinfo {volume} {24}},\
  \bibinfo {pages} {672} (\bibinfo {year} {1962})}\BibitemShut {NoStop}%
\bibitem [{\citenamefont {Hehl}\ and\ \citenamefont {Ni}(1990)}]{Hehl:1990nf}%
  \BibitemOpen
  \bibfield  {author} {\bibinfo {author} {\bibfnamefont {F.~W.}\ \bibnamefont
  {Hehl}}\ and\ \bibinfo {author} {\bibfnamefont {W.-T.}\ \bibnamefont {Ni}},\
  }\href {\doibase 10.1103/PhysRevD.42.2045} {\bibfield  {journal} {\bibinfo
  {journal} {Phys. Rev. D}\ }\textbf {\bibinfo {volume} {42}},\ \bibinfo
  {pages} {2045} (\bibinfo {year} {1990})}\BibitemShut {NoStop}%
\bibitem [{\citenamefont {De~Oliveira}\ and\ \citenamefont
  {Tiomno}(1962)}]{de1962representations}%
  \BibitemOpen
  \bibfield  {author} {\bibinfo {author} {\bibfnamefont {C.}~\bibnamefont
  {De~Oliveira}}\ and\ \bibinfo {author} {\bibfnamefont {J.}~\bibnamefont
  {Tiomno}},\ }\href@noop {} {\bibfield  {journal} {\bibinfo  {journal} {Il
  Nuovo Cimento (1955-1965)}\ }\textbf {\bibinfo {volume} {24}},\ \bibinfo
  {pages} {672} (\bibinfo {year} {1962})}\BibitemShut {NoStop}%
\bibitem [{\citenamefont {Huang}(1994)}]{huang1994dirac}%
  \BibitemOpen
  \bibfield  {author} {\bibinfo {author} {\bibfnamefont {J.~C.}\ \bibnamefont
  {Huang}},\ }\href@noop {} {\bibfield  {journal} {\bibinfo  {journal} {Annalen
  der Physik}\ }\textbf {\bibinfo {volume} {506}},\ \bibinfo {pages} {53}
  (\bibinfo {year} {1994})}\BibitemShut {NoStop}%
\bibitem [{\citenamefont {Papini}(2002)}]{Papini:2002cp}%
  \BibitemOpen
  \bibfield  {author} {\bibinfo {author} {\bibfnamefont {G.}~\bibnamefont
  {Papini}},\ }\href {\doibase 10.1103/PhysRevD.65.077901} {\bibfield
  {journal} {\bibinfo  {journal} {Phys. Rev. D}\ }\textbf {\bibinfo {volume}
  {65}},\ \bibinfo {pages} {077901} (\bibinfo {year} {2002})},\ \Eprint
  {http://arxiv.org/abs/gr-qc/0201098} {arXiv:gr-qc/0201098} \BibitemShut
  {NoStop}%
\bibitem [{\citenamefont {Cai}\ and\ \citenamefont
  {Papini}(1991)}]{Cai:1991wj}%
  \BibitemOpen
  \bibfield  {author} {\bibinfo {author} {\bibfnamefont {Y.~Q.}\ \bibnamefont
  {Cai}}\ and\ \bibinfo {author} {\bibfnamefont {G.}~\bibnamefont {Papini}},\
  }\href {\doibase 10.1103/PhysRevLett.66.1259} {\bibfield  {journal} {\bibinfo
   {journal} {Phys. Rev. Lett.}\ }\textbf {\bibinfo {volume} {66}},\ \bibinfo
  {pages} {1259} (\bibinfo {year} {1991})}\BibitemShut {NoStop}%
\bibitem [{\citenamefont {Ryder}(1998)}]{ryder1998relativistic}%
  \BibitemOpen
  \bibfield  {author} {\bibinfo {author} {\bibfnamefont {L.}~\bibnamefont
  {Ryder}},\ }\href@noop {} {\bibfield  {journal} {\bibinfo  {journal} {Journal
  of Physics A: Mathematical and General}\ }\textbf {\bibinfo {volume} {31}},\
  \bibinfo {pages} {2465} (\bibinfo {year} {1998})}\BibitemShut {NoStop}%
\bibitem [{\citenamefont {Ryder}(2008)}]{ryder2008spin}%
  \BibitemOpen
  \bibfield  {author} {\bibinfo {author} {\bibfnamefont {L.}~\bibnamefont
  {Ryder}},\ }\href@noop {} {\bibfield  {journal} {\bibinfo  {journal} {General
  Relativity and Gravitation}\ }\textbf {\bibinfo {volume} {40}},\ \bibinfo
  {pages} {1111} (\bibinfo {year} {2008})}\BibitemShut {NoStop}%
\bibitem [{\citenamefont {Mashhoon}(1989)}]{mashhoon1989electrodynamics}%
  \BibitemOpen
  \bibfield  {author} {\bibinfo {author} {\bibfnamefont {B.}~\bibnamefont
  {Mashhoon}},\ }\href@noop {} {\bibfield  {journal} {\bibinfo  {journal}
  {Physics Letters A}\ }\textbf {\bibinfo {volume} {139}},\ \bibinfo {pages}
  {103} (\bibinfo {year} {1989})}\BibitemShut {NoStop}%
\bibitem [{\citenamefont {Kapusta}\ \emph {et~al.}(2020)\citenamefont
  {Kapusta}, \citenamefont {Rrapaj},\ and\ \citenamefont
  {Rudaz}}]{Kapusta:2020dco}%
  \BibitemOpen
  \bibfield  {author} {\bibinfo {author} {\bibfnamefont {J.~I.}\ \bibnamefont
  {Kapusta}}, \bibinfo {author} {\bibfnamefont {E.}~\bibnamefont {Rrapaj}}, \
  and\ \bibinfo {author} {\bibfnamefont {S.}~\bibnamefont {Rudaz}},\ }\href
  {\doibase 10.1103/PhysRevD.102.125028} {\bibfield  {journal} {\bibinfo
  {journal} {Phys. Rev. D}\ }\textbf {\bibinfo {volume} {102}},\ \bibinfo
  {pages} {125028} (\bibinfo {year} {2020})},\ \Eprint
  {http://arxiv.org/abs/2009.12010} {arXiv:2009.12010 [hep-th]} \BibitemShut
  {NoStop}%
\bibitem [{\citenamefont {Buzzegoli}\ and\ \citenamefont
  {Kharzeev}(2021)}]{Buzzegoli:2021jeh}%
  \BibitemOpen
  \bibfield  {author} {\bibinfo {author} {\bibfnamefont {M.}~\bibnamefont
  {Buzzegoli}}\ and\ \bibinfo {author} {\bibfnamefont {D.~E.}\ \bibnamefont
  {Kharzeev}},\ }\href {\doibase 10.1103/PhysRevD.103.116005} {\bibfield
  {journal} {\bibinfo  {journal} {Phys. Rev. D}\ }\textbf {\bibinfo {volume}
  {103}},\ \bibinfo {pages} {116005} (\bibinfo {year} {2021})},\ \Eprint
  {http://arxiv.org/abs/2102.01676} {arXiv:2102.01676 [hep-th]} \BibitemShut
  {NoStop}%
\bibitem [{\citenamefont {Balitsky}\ and\ \citenamefont
  {Ji}(1997)}]{Balitsky:1997rs}%
  \BibitemOpen
  \bibfield  {author} {\bibinfo {author} {\bibfnamefont {I.}~\bibnamefont
  {Balitsky}}\ and\ \bibinfo {author} {\bibfnamefont {X.-D.}\ \bibnamefont
  {Ji}},\ }\href {\doibase 10.1103/PhysRevLett.79.1225} {\bibfield  {journal}
  {\bibinfo  {journal} {Phys. Rev. Lett.}\ }\textbf {\bibinfo {volume} {79}},\
  \bibinfo {pages} {1225} (\bibinfo {year} {1997})},\ \Eprint
  {http://arxiv.org/abs/hep-ph/9702277} {arXiv:hep-ph/9702277} \BibitemShut
  {NoStop}%
\bibitem [{\citenamefont {Shifman}\ \emph {et~al.}(1979)\citenamefont
  {Shifman}, \citenamefont {Vainshtein},\ and\ \citenamefont
  {Zakharov}}]{Shifman:1978bx}%
  \BibitemOpen
  \bibfield  {author} {\bibinfo {author} {\bibfnamefont {M.~A.}\ \bibnamefont
  {Shifman}}, \bibinfo {author} {\bibfnamefont {A.~I.}\ \bibnamefont
  {Vainshtein}}, \ and\ \bibinfo {author} {\bibfnamefont {V.~I.}\ \bibnamefont
  {Zakharov}},\ }\href {\doibase 10.1016/0550-3213(79)90022-1} {\bibfield
  {journal} {\bibinfo  {journal} {Nucl. Phys. B}\ }\textbf {\bibinfo {volume}
  {147}},\ \bibinfo {pages} {385} (\bibinfo {year} {1979})}\BibitemShut
  {NoStop}%
\bibitem [{\citenamefont {Ji}(1997)}]{Ji:1996ek}%
  \BibitemOpen
  \bibfield  {author} {\bibinfo {author} {\bibfnamefont {X.-D.}\ \bibnamefont
  {Ji}},\ }\href {\doibase 10.1103/PhysRevLett.78.610} {\bibfield  {journal}
  {\bibinfo  {journal} {Phys. Rev. Lett.}\ }\textbf {\bibinfo {volume} {78}},\
  \bibinfo {pages} {610} (\bibinfo {year} {1997})},\ \Eprint
  {http://arxiv.org/abs/hep-ph/9603249} {arXiv:hep-ph/9603249} \BibitemShut
  {NoStop}%
\bibitem [{\citenamefont {Reinders}\ \emph {et~al.}(1985)\citenamefont
  {Reinders}, \citenamefont {Rubinstein},\ and\ \citenamefont
  {Yazaki}}]{Reinders:1984sr}%
  \BibitemOpen
  \bibfield  {author} {\bibinfo {author} {\bibfnamefont {L.~J.}\ \bibnamefont
  {Reinders}}, \bibinfo {author} {\bibfnamefont {H.}~\bibnamefont
  {Rubinstein}}, \ and\ \bibinfo {author} {\bibfnamefont {S.}~\bibnamefont
  {Yazaki}},\ }\href {\doibase 10.1016/0370-1573(85)90065-1} {\bibfield
  {journal} {\bibinfo  {journal} {Phys. Rept.}\ }\textbf {\bibinfo {volume}
  {127}},\ \bibinfo {pages} {1} (\bibinfo {year} {1985})}\BibitemShut {NoStop}%
\bibitem [{\citenamefont {Morita}\ and\ \citenamefont
  {Lee}(2010)}]{Morita:2009qk}%
  \BibitemOpen
  \bibfield  {author} {\bibinfo {author} {\bibfnamefont {K.}~\bibnamefont
  {Morita}}\ and\ \bibinfo {author} {\bibfnamefont {S.~H.}\ \bibnamefont
  {Lee}},\ }\href {\doibase 10.1103/PhysRevD.82.054008} {\bibfield  {journal}
  {\bibinfo  {journal} {Phys. Rev. D}\ }\textbf {\bibinfo {volume} {82}},\
  \bibinfo {pages} {054008} (\bibinfo {year} {2010})},\ \Eprint
  {http://arxiv.org/abs/0908.2856} {arXiv:0908.2856 [hep-ph]} \BibitemShut
  {NoStop}%
\bibitem [{\citenamefont {Gubler}\ and\ \citenamefont
  {Satow}(2019)}]{Gubler:2018ctz}%
  \BibitemOpen
  \bibfield  {author} {\bibinfo {author} {\bibfnamefont {P.}~\bibnamefont
  {Gubler}}\ and\ \bibinfo {author} {\bibfnamefont {D.}~\bibnamefont {Satow}},\
  }\href {\doibase 10.1016/j.ppnp.2019.02.005} {\bibfield  {journal} {\bibinfo
  {journal} {Prog. Part. Nucl. Phys.}\ }\textbf {\bibinfo {volume} {106}},\
  \bibinfo {pages} {1} (\bibinfo {year} {2019})},\ \Eprint
  {http://arxiv.org/abs/1812.00385} {arXiv:1812.00385 [hep-ph]} \BibitemShut
  {NoStop}%
\bibitem [{\citenamefont {Ashman}\ \emph {et~al.}(1988)\citenamefont {Ashman}
  \emph {et~al.}}]{EuropeanMuon:1987isl}%
  \BibitemOpen
  \bibfield  {author} {\bibinfo {author} {\bibfnamefont {J.}~\bibnamefont
  {Ashman}} \emph {et~al.} (\bibinfo {collaboration} {European Muon}),\ }\href
  {\doibase 10.1016/0370-2693(88)91523-7} {\bibfield  {journal} {\bibinfo
  {journal} {Phys. Lett. B}\ }\textbf {\bibinfo {volume} {206}},\ \bibinfo
  {pages} {364} (\bibinfo {year} {1988})}\BibitemShut {NoStop}%
\end{thebibliography}%

\end{document}